\UseRawInputEncoding
\documentclass[10pt,twocolumn]{article}

\usepackage[
    top=0.5in,
    bottom=0.5in,
    left=0.5in,
    right=0.5in,
    columnsep=0.25in
]{geometry}

\usepackage{graphicx}
\usepackage{authblk}
\usepackage{amsfonts,amsmath,amssymb,amsthm}
\usepackage{epstopdf}
\usepackage{upgreek,xspace}
\usepackage{chngcntr}
\usepackage{hyperref}
\hypersetup{colorlinks,allcolors=blue}
\usepackage[version=3]{mhchem}
\usepackage{soul}

\usepackage{setspace}

\newcommand{\be}{\begin{equation}}
\newcommand{\ee}{\end{equation}}

\newcommand{\Tg}{$T_\mathrm{g}$}
\newcommand{\Ta}{$T_\mathrm{anneal}$}
\newcommand{\Po}{$P_\mathrm{O_2}$}
\newcommand{\AO}{Al$_2$O$_3$}

\newcommand{\dC}{$^\circ$C}

\newcommand{\AVO}{AlV$_2$O$_4$}

\newcommand{\Pog}{g-$P_\mathrm{O_2}$}
\newcommand{\Poa}{a-$P_\mathrm{O_2}$}

\title{AlV$_2$O$_4$ thin films via \textit{in-situ} interfacial topotaxy}

\author[1,2]{Jeong~Rae~Kim}
\author[1,2]{Alexis~Ashby}
\author[1,2]{Sandra~Glotzer}
\author[3]{Darryl~Shima}
\author[3]{Ganesh~Balakrishnan}
\author[1,2]{Joseph~Falson\thanks{Email: falson@caltech.edu}}

\affil[1]{Department of Applied Physics and Materials Science,
California Institute of Technology, Pasadena, California 91125, USA}
\affil[2]{Institute for Quantum Information and Matter,
California Institute of Technology, Pasadena, California 91125, USA}
\affil[3]{Center for High Technology Materials,
University of New Mexico, Albuquerque, New Mexico 87106, USA}

\date{}

\begin{document}
\maketitle
\begin{abstract}
Conventional oxide epitaxy approaches face challenges when the oxidation conditions of constituent elements differ significantly. Here we demonstrate that V--O thin films can serve as solid-phase precursors for epitaxial \AVO, comprising a pyrochlore V$^{2.5+}$ network coexisting with AlO$_4$ tetrahedra within the spinel structure. The epitaxial \AVO/\AO~(0001) heterostructures are realized via interfacial topotactic transformation, involving
V--O growth at a moderate temperature followed by \textit{in-situ}~ultra-high-temperature post-annealing to drive a reaction with the \AO~substrate. Transmission electron microscopy and temperature-dependent X-ray diffraction analyses reveal excellent structural characteristics that closely reproduce the known charge-ordering transition. This study presents a novel approach to realizing epitaxial structures with convoluted oxidation states where thermodynamic and kinetic barriers would otherwise limit synthesizability.
\end{abstract}

\section*{Introduction}
Thin-film science is a rich field where interfacially stabilized artificial heterostructures host a complex array of physical phenomena~\cite{ohring2002materials,herman2013epitaxy}. The synthesis of a thin film typically begins by delivering the constituent elements of the target material to a substrate under suitable thermokinetic conditions. The delivery method generally involves physical or chemical evaporation of the source materials onto a template substrate to access a synthesis window conducive to forming the desired crystal phase and microstructure~\cite{mattox2010handbook,hitchman1993chemical}. However, achieving high-quality thin films with precise stoichiometry can occasionally be hindered by compositional chemistry and method-specific limitations. A less commonly employed approach is to utilize the substrate itself as the source material. In this process, the delivered material undergoes a chemical reaction with the substrate components at the interface, forming a thin-film layer. The substrate therefore offers additional functionality in the growth process beyond its typical role of templating growth. The interfacial reaction phenomenon is often difficult to predict and, in fact, is regarded as a source of defects in most deposition cases. However, when the necessary conditions are met, such as in the model SiO$_x$/Si system~\cite{deal1965general}, the interfacial reaction of a uniform thin-film layer can be precisely controlled. Furthermore, if the target material and the substrate are in coherent crystallographic relation, it is even possible to fabricate fully epitaxial heterostructures through an interfacial topotactic transformation in a process termed ``interfacial topotaxy" \cite{Shannon1964,tung1983formation}. Therefore, we propose that interfacial reaction should rather be considered as an accessible tool for crafting a diverse array of materials, especially in the thermal limit. 

For semiconductor substrates with relatively low melting points, the required temperature for promoting interfacial reactions is often within the range of regular thin-film resistive heating technologies~\cite{tung1983formation,strohbeen2024molecular}. However, this does not extend to refractory ceramics, including oxides, where \textit{in-situ} interfacial reactions are constrained by temperature limitations as significant thermal activation of the substrate and film materials is required. In previous studies of oxide films, interfacial topotaxy has predominantly been achieved via solid state reactions~\cite{hesse1981formation,hesse1983solid}. High-temperature, \textit{ex-situ} post-annealing was often required for the deposited and substrate materials to form the film layer, typically in an ambient atmosphere~\cite{shy2005characterization,huon2022solid}. Consequently, the interfacial topotaxy of oxides has largely been restricted to compounds with common oxidation states~\cite{sun2004topotaxial,hesse2021interfacial}.

In this work, we demonstrate the utility of \textit{in-situ} interfacial reactions in the realization of epitaxial \AVO/\AO~(0001). \AVO~exhibits a rare 2.5+ average oxidation state of vanadium and has not been synthesized as a thin film to date. Our method involves a two-step process consisting of initial deposition of vanadium oxide on an \AO~substrate, followed by high temperature annealing under tightly controlled oxygen conditions. This is performed \textit{in-situ} using a laser heating method previously employed to synthesize early transition metal oxides~\cite{kim2025high,kim2025superconducting,glotzer2026thermally,birkholzer2026synthesis}. Mass transfer across the heterointerface results in the formation of \AVO, which is otherwise not possible to synthesize in a single-step process. This study reveals that precision control over interfacial topotaxy can be utilized as a novel synthesis knob to realize suboxides with difficult-to-control oxidation states.

\section*{Results and discussion}
\subsection*{Overview of growth}
The formation of complex crystals allows metal ions to adopt unconventional charge states, such as mixed valences, charge disproportionation, and charge density waves~\cite{khomskii2014transition,cox1976crystal,johnston2014charge}. A diverse array of quantum phases, particularly superconductivity, are investigated in close connection with these charge instabilities~\cite{lee2006doping,kim2022superconductivity}, while material searches increasingly target compounds exhibiting progressively rarer charge states. Vanadium-based materials have played a pivotal role in this field. The numerous stable oxidation states --- +0, +2, +3, +4, and +5 --- give rise to a wide range of quantum materials, spanning from intermetallics to ionic compounds~\cite{wilson2024v3sb5,huang2024non,kondo1997liv,katayama2009anomalous,yun2025direct}. Among them, the spinel \AVO~is unique in possessing an average +2.5 valence state~\cite{matsuno2001charge,horibe2006spontaneous,browne2017persistent}. Below 700~K, \AVO~undergoes a charge-ordering transition in which the nominal V$^{2.5+}$ state undergoes charge disproportionation, and this transition is accompanied by electrical and magnetic anomalies. Leveraging the tunability of its charge order through doping and pressure~\cite{matsuno2003charge,horibe2005doping,kalavathi2013pressure}, \AVO~emerges as an excellent material platform for studying the interplay of multivalency, geometric frustration, and strong electron correlations.

The synthesis of \AVO~thin films is complicated by the distinct oxidation kinetics of Al and V. To review the thin-film synthesis of the V--O system, V$^{2+}$ and V$^{3+}$ are relatively rare oxidation states and thus syntheses of VO and V$_2$O$_3$ thin films have been conducted in high vacuum conditions~\cite{rata2004growth,rata2005strain,brockman2011increased,brockman2012substrate}. The oxygen stoichiometry of V--O films is highly sensitive to the background oxygen partial pressure, necessitating precise control of oxygen dosing for optimal film quality. And while fully oxidized \AO~exhibits excellent stability over a wide range of thermodynamic environments including high-temperature high-vacuum conditions~\cite{shang2024ellingham}, the Al$_2$O suboxide is highly volatile and forms during the oxidation process from Al to \AO. Thus, previous reports on \AO~epitaxy operate at high pressures exceeding 1.0$\times$10$^{-3}$ Torr in order to suppress material loss at high temperatures~\cite{majer2024adsorption}. Collectively, this implies that no viable thin-film thermodynamic conditions exist to simultaneously accommodate V$^{2.5+}$ and suppress Al$_2$O evaporation. Such cases are not encountered in the synthesis of bulk AlV$_2$O$_4$ performed in sealed containers where material evaporation is suppressed~\cite{matsuno2001charge,browne2017persistent}. These constraints prompt us to explore interfacial reaction as a means to extract fully oxidized \AO~from the substrate to compose \AVO. The core of the strategy lies in optimizing the V$^{2.5+}$ oxidation state under thermal conditions sufficient for the AlV$_2$O$_4$ formation.

\begin{figure}[t]
  \centering
   \includegraphics[width=85mm]{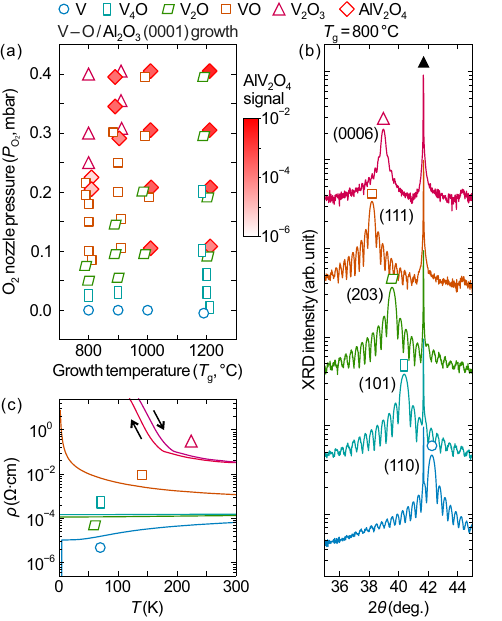}
   \caption{Standard growth of V--O on \AO~(0001). (a) Growth phase diagram in the~\Tg-\Po~parameter space with symbols corresponding to V (blue circle), V$_4$O (cyan rectangle), V$_2$O (green parallelogram), VO (orange square), V$_2$O$_3$ (magenta triangle), and \AVO~(red diamond). The \AVO~signal was defined as the \AVO~(111)$_{pc}$/\AO~(0006) XRD intensity ratio. (b,c) XRD 2$\theta$--$\theta$ (b) and temperature-dependent resistivity (c) measurements of the series of single-phase V--O films grown at fixed \Tg~= 800 \dC~and varying \Po~= 0, 0.025, 0.05, 0.15, and 0.3 mbar. The XRD signals from the \AO~substrates are indicated by black triangles. The resistivity of V$_2$O$_3$ was measured during both cooling and heating, with arrows indicating the direction of the temperature sweep.}
   \label{Fig1}
\end{figure}

\begin{figure*}[t]
  \centering
   \includegraphics[width=170mm]{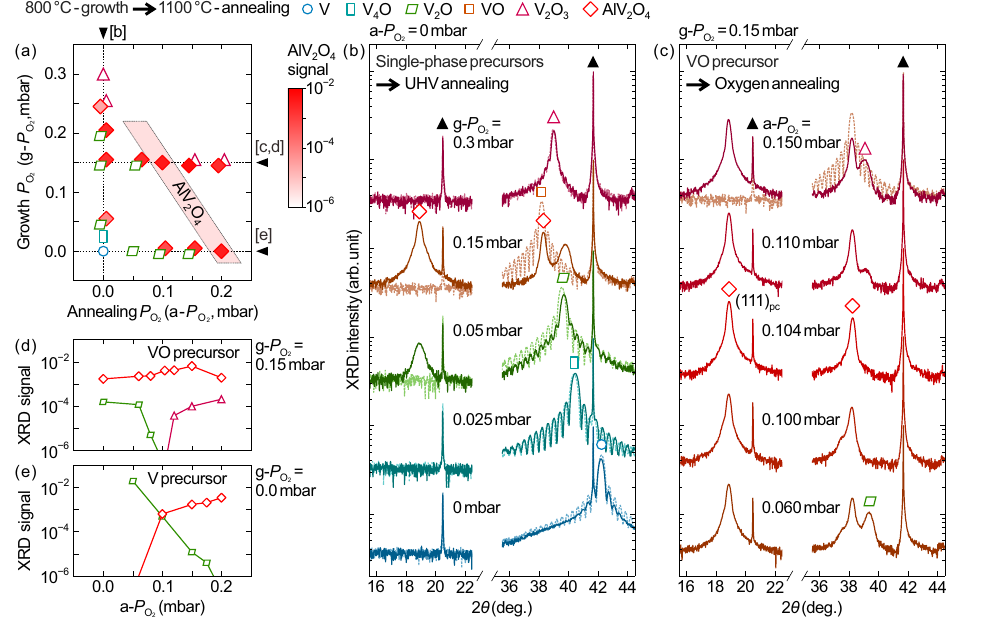}
   \caption{Two-step processing methods to achieve single-phase \AVO~films. (a) Synthesis phase diagram in the \Poa-\Pog~parameter space. The growth (annealing) temperature is fixed at 800~\dC~(1100~\dC). (b) Test of spontaneous reactivity between \AO~substrate and V, V$_4$O, V$_2$O, VO, and V$_2$O$_3$. XRD 2$\theta$--$\theta$ plots (solid lines) of samples corresponding to (\Poa,~\Pog) = (0, 0), (0, 0.025), (0, 0.05), (0, 0.15), and (0, 0.3) points in (a). (c) Synthesis optimization of \AVO~phase with an initial VO/\AO~(0001) precursor film. XRD 2$\theta$--$\theta$ plots (solid lines) of samples corresponding to (\Poa,~\Pog) = (0.06, 0.15), (0.1, 0.15), (0.104, 0.15), (0.11, 0.15), and (0.15, 0.15) points in (a). The plots of precursors corresponding to each case are drawn with dashed lines in the background (b,c). (d,e) Comparison of the VO (d) and V (e) precursors. For the \Pog~= 0 and 0.15 mbar lines, \AO~(0006)-normalized XRD intensities of \AVO, V$_2$O, and V$_2$O$_3$ phases are plotted as a function of \Poa.}
   \label{Fig2}
\end{figure*}

\subsection*{Standard growth of thin-film V--O/\AO~(0001) system}

We first examine the standard growth phase diagram of V--O/\AO~(0001) thin films. The growth temperature (\Tg) was varied over the \Tg~= 800-1200 \dC~range. Elemental vanadium and molecular oxygen were supplied by an e-beam evaporator and a custom gas injector system. For all growths, the vanadium flux was kept constant, while the oxygen flux quantified by the local nozzle pressure (\Po) was varied. Figure \ref{Fig1}a shows the \Tg--\Po~diagram in which we identify six different phases. The bulk V--O phase diagram has been extensively studied, and most of the suboxides are non-stoichiometric~\cite{wriedt1989ov}. The phases for which a consensus on crystal structure has been reached are typically designated as $\alpha$ (body-centered cubic, $Im\overline{3}m$), $\beta$ (body-centered tetragonal, $I4/mmm$), $\gamma$ (monoclinic, $P2_{1}/c$), and $\delta$ (rock-salt, $Fm\overline{3}m$) phases, starting from the lowest oxidation state~\cite{henry1970vanadium,davydov2009ordered,banus1972electrical}. In this article, they are designated as V, V$_4$O, V$_2$O, and VO, according to their representative oxidation states. The stoichiometry and crystal structure of V$_2$O$_3$ are relatively well established~\cite{dernier1970crystal}, and it exhibits the highest vanadium oxidation state among the phases considered. In addition, the target \AVO~phase is detected.

The five thin-film phases of vanadium oxide are identified by their structural and electrical properties (Fig.\ref{Fig1}b,c). Figure \ref{Fig1}b shows the X-ray diffraction (XRD) analysis of the series of single-phase films grown at fixed \Tg~= 800 \dC~and varying \Po~= 0, 0.025, 0.05, 0.15, and 0.3 mbar. They are clearly distinguishable based not only on out-of-plane lattice spacing but also by  reciprocal space mapping (RSM) analysis (Fig.S1 in the
Supporting Information). We have also obtained temperature-dependent resistivity measurements of films to verify their character, as shown in Fig.\ref{Fig1}c. The characteristics of the V, VO, and V$_2$O$_3$ films are consistent with those of the previously reported bulk crystals and thin films~\cite{banus1972electrical,rata2004growth,mcwhan1973metal,brockman2011increased}. Studies on V$_4$O and V$_2$O are limited and to date, no single-phase thin films have been reported. In this work, V$_4$O and V$_2$O exhibit conventional metallic behavior, with the resistivity of V$_4$O being slightly higher than that of V$_2$O. With this exception, the general trend is that resistivity rises as the oxidation state of V increases.

Analyzing the growth phase diagram (Fig.\ref{Fig1}a) allows us to identify two key points. First, with the \AVO~phase excluded, the thin-film V--O system closely follows the expected thermodynamic trends; higher vanadium oxidation states are obtained at lower \Tg~and higher \Po. We previously reported an unusual oxygen-diffusion-controlled growth mode of the thin-film Ti--O/\AO~(0001) system~\cite{kim2025high}. In that case, oxygen diffusion was highly active and the diffusion-controlled regime, characterized by a reversal of the thermodynamic temperature dependence, was observed over a wide range of the growth phase diagram. In the case of V--O/\AO~(0001), oxygen diffusion is observed only to a limited extent near the (\Tg, \Po)~= (1200 \dC, 0 mbar) point, presumably due to the higher electronegativity of vanadium. Second, the degree of the \AVO~formation is strongly dependent on \Tg. At relatively low \Tg~= 800, 900 \dC, a small amount of \AVO~emerged under pressure conditions intermediate between those optimal for VO and V$_2$O$_3$. As \Tg~rises, both the degree of the interfacial reaction and the \Po~window increase significantly.

The outcome of this phase diagram is that a standard V--O/\AO~(0001) growth process cannot produce single-phase AlV$_2$O$_4$ thin films; at low temperatures, the extent of the interfacial reaction is insufficient, whereas at high temperatures, the V$^{2.5+}$ oxidation state could not be achieved.

\subsection*{Two-step processing of phase pure AlV$_2$O$_4$ films}

We now implement a two-step process to simultaneously achieve the desired degree of interfacial reaction and sufficient oxidation. The thin-film  V--O system functions as a solid-phase precursor in this approach. It consists of (1) standard V--O growth at \Tg~= 800~$^\circ$C, followed by (2) an \textit{in-situ} post-annealing process at \Ta~= 1100~$^\circ$C. Here, the \Po~for the growth (\Pog) and annealing (\Poa) steps are individually controlled, and the resulting~(\Pog,\Poa)--parameter space is refined in Fig.\ref{Fig2}(a). The \Tg~=~800 and \Ta~=~1100~$^\circ$C values were selected by considering a range of factors, including the accessible oxidation window, degree of interfacial reaction, thin-film uniformity, growth rate, and material desorption at elevated temperatures.

As an entry point, we evaluate the spontaneous reactivity of five single-phase vanadium oxide thin-film precursors with \AO~substrates by ultra high vacuum (UHV) post-annealing (\Poa~= 0 mbar). As we discuss below, the 1100~$^\circ$C value is sufficient to trigger interfacial reaction but is insufficient to induce thermodynamic oxygen reduction or kinetic oxygen diffusion. Figure \ref{Fig2}(b) presents the XRD plots of the single-phase V, V$_4$O, V$_2$O, VO, and V$_2$O$_3$ samples before (dotted lines, same samples shown in Fig.\ref{Fig1}(b)) and after (solid lines) the UHV annealing. Unlike the five vanadium oxide phases, only \AVO~exhibits a strong peak near 2$\theta$~=~19$^\circ$, making it a reliable indicator for quantifying the synthesis of \AVO. Given the uncertainty in determining the crystal phase and orientation of spinel \AVO, we assume the pseudocubic (pc) crystal structure of \AVO~and designate this peak as \AVO~(111)$_{pc}$ throughout this study, unless otherwise noted. The \AVO~signal is defined as the \AVO~(111)$_{pc}$/\AO~(0006) XRD intensity ratio. The UHV annealing results in pronounced \AVO~formation from the VO precursor, in addition to V$_2$O phase formation, with comparatively less \AVO~forming in the case of the V$_2$O precursor. This can be attributed to elemental conservation; a portion of the initial thin-film layer with V$^{2+}$ or V$^{1+}$ oxidation states reacts to form the V$^{2.5+}$ \AVO~layer. If the oxygen content of the layer is conserved, the remaining unreacted portion undegoes reduction to a lower oxidation state (V$^{1+}$) than the initial state. The V, V$_4$O, and V$_2$O$_3$ precursors do not exhibit spontaneous reaction with \AO. V$_4$O and V$_2$O$_3$ layers remain essentially unaffected by the UHV annealing, whereas the V layer exhibits a slight peak shift due to oxygen diffusion from the substrate.

To achieve single-phase \AVO~thin films, we refine the \textit{in-situ} post-annealing step by incorporating precise \Poa~control. The VO precursor is selected as the initial state (\Tg~= 800~$^\circ$C, \Pog~= 0.15 mbar) for its optimal reactivity to form \AVO. Figure \ref{Fig2}(c) shows the XRD plots (solid lines) of such postannealed samples (\Pog~= 0.15 mbar and \Poa~= 0.06, 0.1, 0.104, 0.11, and 0.15 mbar). For visual clarity, only one dotted plot is shown in Fig.\ref{Fig2}(c) as the initial state for all samples is the same (VO). This series reveals a narrow \Poa-window where single-phase \AVO~films are realized. At \Poa~= 0.104~mbar, \AVO~is the sole structure in XRD, illustrating the growth of a phase-pure film (this sample corresponds to the (\Pog, \Poa) = (0.15 mbar, 0.104 mbar) point in Fig.\ref{Fig2}(a)). Under- and over-oxidation result in the formation of V$_2$O and V$_2$O$_3$ byproducts.

We lastly discuss the choice of the precursor material. Evidently, V$_2$O$_3$ is disqualified as it shows no evidence of the \AVO~formation or oxygen reduction below 1200~$^\circ$C, while above, it exhibited loss of the layer itself due to thermal desorption/evaporation. The V, V$_4$O, V$_2$O, and VO could all be viable precursor phases for the synthesis of \AVO~thin films. The cases of the V and VO precursors are highlighted in Fig.\ref{Fig2}(d,e), where two line cuts in Fig.\ref{Fig2}(a) (\Pog~= 0 and 0.15 mbar) are compared. Figure \ref{Fig2}(d) quantitatively illustrates the \Poa-optimization process using the VO precursor discussed above. While the \AVO~signal remains largely invariant, the signals from the V$_2$O and V$_2$O$_3$ byproducts strongly depend on the \Poa. Using the V precursor, the single-phase \AVO~is obtained at a different optimal \Poa~of 0.2 mbar (Fig.\ref{Fig2}(e) and Fig.S2 in the Supporting Information). In this case, the change in the V$_2$O signal with respect to \Poa~is relatively gradual, signifying that the interfacial reaction proceeds only after the supplied oxygen generates a finite amount of V$_2$O species capable of reacting with \AO. If the film layer is substantially thicker or the thermal energy available for the interfacial reaction is insufficient, the V precursor may result in a non-uniform compositional depth profile.

\begin{figure}[t]
  \centering
   \includegraphics[width=85mm]{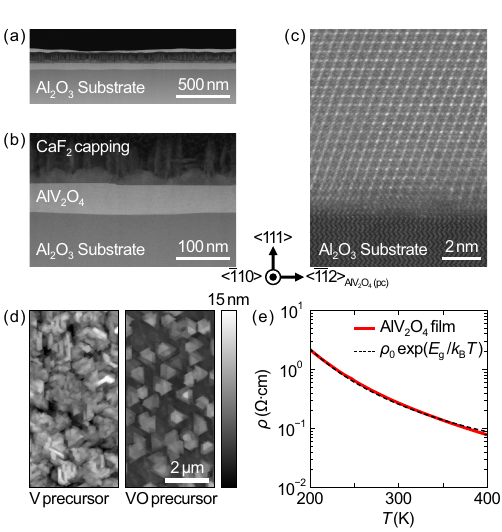}
   \caption{Structural and electrical characterization of \AVO~films. (a--c) High-angle annular dark-field STEM (HAADF-STEM) image of a CaF$_2$/\AVO/\AO~(0001) sample, taken from the \AO~[1$\overline{1}$00] zone axis. (d) Atomic force microscopy images of \AVO~film prepared from the V and VO precursors. (e) Temperature-dependent resistivity measurement of the \AVO~film and a $\rho$~=~$\rho_{0}\cdot$exp($E_{g}$/$k_{B}T$) fitting ($\rho_{0}$~$\approx$~3.5~m$\Omega\cdot$cm, $E_{g}$~$\approx$~0.11 eV). The VO precursors were used for (a--c, e).}
   \label{Fig3}
\end{figure}

\subsection*{Properties of epitaxial AlV$_2$O$_4$~films}
Having narrowed down the synthesis pathway to phase-pure \AVO~films using a two-step process, we now study their physical properties. In Fig.\ref{Fig3}(a-c) we present cross-sectional scanning transmission electron microscopy (STEM) characterization of a single-phase \AVO~sample along the \AO~[1$\overline{1}$00] zone axis. The \AVO~sample is prepared from a VO precursor and capped with a CaF$_2$~layer (discussed below). Despite the substantial mass transfer taking place, the heterointerface remains uniform and straight over several micrometers, with local disorder on the $\approx$1~nm length scale. The high-resolution STEM image given in Fig.\ref{Fig3}(c) reveals the characteristic pyrochlore sublattice of V in the spinel structure. The \AVO~film layer is epitaxial to the \AO~substrate, with the relationship of \AVO~[11$\overline{2}$]~$\parallel$~\AO~[11$\overline{2}$0]. We also observe twin domains in the (111)-oriented cubic lattice, similar to those in NbO/\AO~(0001) thin films~\cite{kim2025superconducting}. Spatially resolved STEM energy dispersive X-ray spectroscopy analysis is given in Fig.S3 in the Supporting Information. The two-step process facilitates interfacial reaction (bottom-up) and surface oxidation (top-down), potentially leading to structural or chemical anisotropy along the out-of-plane direction. Despite this, the \AVO~thin film exhibits a uniform distribution of Al, V, and O while maintaining a sharp interface with the \AO~substrate.

Figure \ref{Fig3}(d) show the surface morphology of $\sim$~50 nm-thick \AVO/\AO~(0001) films, measured using atomic force microscopy. We compare two single-phase \AVO~films from V and VO precursors. Despite their similar X-ray characteristics, the sample from the VO precursor exhibits superior macroscopic surface smoothness. The few-nanometer-high triangular patches can be associated with the (111)-oriented cubic lattice, as well as the surface step observed in Fig.\ref{Fig3}(b).

Finally, we report the electrical properties of the single-phase \AVO~films. Bulk \AVO~is a semiconductor and exhibits a resistivity anomaly associated with a charge ordering phase transition near 700 K (discussed below)~\cite{matsuno2001charge,matsuno2003charge}. The bulk semiconducting state at $T~>$ 700 K ($<$ 700 K) has an activation energy of $\sim$0.26 eV (0.39 eV). Our temperature-dependent resistivity measurements of the \AVO~sample from a VO precursor also reveal semiconducting behavior over the 200-400~K temperature range with $E_{g}$~=~0.11 eV. At 400 K, the resistivity is approximately two orders of magnitude lower than that reported for bulk \AVO~(10$\sim$20~$\Omega\cdot$cm)~\cite{matsuno2001charge}.

\begin{figure}[t]
  \centering
   \includegraphics[width=85mm]{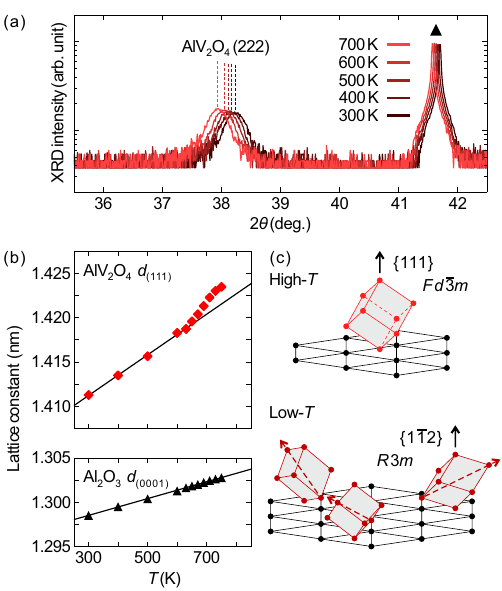}
   \caption{Temperature-driven structural transition of a \AVO~film. (a) XRD 2$\theta$--$\theta$ plots of a \AVO/\AO~(0001) sample measured at $T$~= 300, 400, 500, 600, and 700 K. (b) Plots of out-of-plane lattice spacing $d$~of the \AVO~film layer and \AO~substrate as a function of $T$. The linear fits for both are overlaid. (c) Schematic illustrations of the crystallographic configurations of $Fd\overline{3}m$- and $R3m$-\AVO~on the \AO~(0001) surface lattices.}
   \label{Fig4}
\end{figure}

\begin{figure}[t]
  \centering
   \includegraphics[width=85mm]{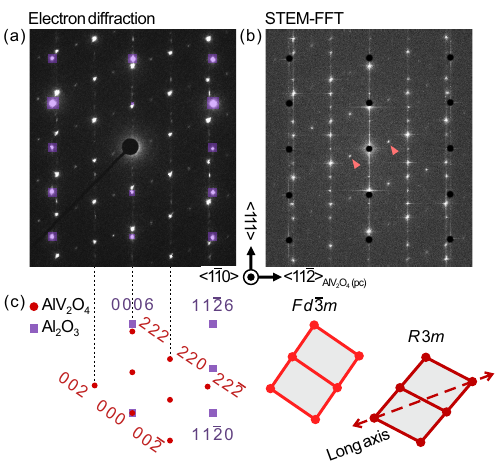}
   \caption{Superlattice structures the \AVO~film. (a,b) TEM electron diffraction pattern (a) and a fast fourier transform of HAADF-STEM (b) taken from the \AO~[1$\overline{1}$00] zone axis. The signals from the \AO~substrate were masked by purple squares (a) and black circles (b). (c) Schematic illustrations of the reciprocal lattices of an \AVO/\AO~(0001) heterostructure and the real \AVO~lattices ($Fd\overline{3}m$ and $R3m$). The (1/2, 1/2, 1/2)$_{pc}$ superlattice peaks are indicated by red arrows in (b).}
   \label{Fig5}
\end{figure}

\subsection*{Charge-ordering transition of epitaxial AlV$_2$O$_4$~films}

The charge-ordering transition is a distinctive characteristic of \AVO~\cite{matsuno2001charge,horibe2006spontaneous,browne2017persistent}. According to previous bulk studies, above the transition temperature (\textit{T} $>$ 700 K)~\cite{matsuno2001charge}, \AVO~adopts a homogeneous V$^{2.5+}$ charge distribution and a regular cubic spinel structure ($Fd\overline{3}m$). The transition induces electron redistribution, leading to the coexistence of higher- and lower-valence V sites and a concomitant rhombohedral distortion. The $R3m$ rhombohedral spinel structure develops one elongated and three shortened \AVO~(111)$_{pc}$ axes. We probe the charge state of epitaxial \AVO~films by examining their associated structural changes through temperature-dependent XRD and electron diffraction.

Figure~\ref{Fig4} illustrates the temperature-dependent XRD experiments on a single phase \AVO~thin film. The measurements have been performed on a tunable temperature sample stage under a background pressure $<$ 3.0$\times$10$^{-1}$ mbar. We found that the \AVO~layer undergoes oxygen-related degradation under these conditions. Therefore, in order to preserve the film under these conditions, here we have introduced a $\sim$40-nm thick CaF$_2$ capping layer that was deposited at room temperature in UHV \textit{in situ}. The CaF$_2$~layer is partially crystalline near the CaF$_2$/\AVO~interface (Fig.\ref{Fig3}(b)), while the rest is amorphous. Figure \ref{Fig4}(a) shows the XRD 2$\theta-\theta$ plots at discrete constant temperatures. Both \AO~(0006) and \AVO~(111)$_{pc}$ peaks shift towards lower angles with increasing temperature, reflecting the thermal expansion of the lattice. Lattice constant values were extracted and plotted in Fig.\ref{Fig4}(b). The \AO~(0001) $d$ shows a near-linear thermal expansion, as expected over this temperature range~\cite{chikh2016situ}. The \AVO~(111)$_{pc}$ $d$ exhibits similar behavior at low temperatures but shows a deviation from linear expansion near 700~K, which is close in temperature to the known transition of bulk \AVO. The observed lattice dynamics indicate that the rhombohedral short axis of the as-grown epitaxial \AVO~film is oriented along the out-of-plane \AO~(0001) direction. Because the heterostructure has threefold symmetry, triply degenerate rhombohedral domains are expected to emerge upon cooling from the initially cubic \AVO~layer synthesized at \Ta~= 1100~\dC, as conceptually illustrated in Fig.\ref{Fig4}(c). 

Superlattice structure analysis has provided valuable insight into the nature of the charge-ordered state in bulk \AVO~crystals~\cite{matsuno2001charge,horibe2006spontaneous,browne2017persistent,okawa2024charge,talanov2018vanadium,radaelli2005orbital}. In particular, (1/2, 1/2, 1/2)$_{pc}$-type lattice doubling along the long axis of $R3m$-\AVO~has been known since the earliest studies. We inspect the electron diffraction patterns to determine whether the superlattice structure is reproduced in the epitaxial \AVO~films. The same transmission electron microscopy geometry as in the previous section is adopted. The corresponding result is shown in Fig.\ref{Fig5}(a). A fast Fourier transform (FFT) of the HAADF-STEM image is shown for enhanced visual clarity. (Fig.\ref{Fig5}(b)). The vertical direction corresponds to the out-of-plane \AO~(0001), and interfering \AO~peaks were masked to selectively display the \AVO~signals. Here, we detect the same (1/2, 1/2, 1/2)$_{pc}$ superlattice spots from the epitaxial \AVO~film layer. Its propagation direction was tilted by $\sim$70$^\circ$ relative to the out-of-plane direction. Together with the preceding temperature-dependent XRD analysis, the electron diffraction measurement establishes the structure and orientation of the epitaxial $R3m$~(\AVO) thin film. 

We also note that, in addition to the (1/2, 1/2, 1/2)$_{pc}$~spots, we observe unknown (1/3, 1/3, 1/3)$_{pc}$~spots propagating towards the out-of-plane direction. Various models for the \AVO~charge ordering have been proposed to date.
The pyrochlore V sublattice consists of alternating kagome and triangular planes along a \AVO~(111)$_{pc}$ direction [V$_3$--V--V$_3$--V--...]. The proposed models include three-one type ordering [V$_{3}^{(7.5-3\delta)+}$--V$^{(2.5+3\delta)+}$--...]~\cite{matsuno2001charge}, heptamer clustering [V$_{7}^{17+}$--V$^{3+}$--...]~\cite{horibe2006spontaneous}, and coupled trimer-tetramer clustering [V$_{4}^{8+}$--V$_{3}^{9+}$--V$^{3+}$--...]~\cite{browne2017persistent}. Our finding suggests the possibility of a new type of charge-ordering state in the epitaxial \AVO~thin films. Future studies will focus on the exact charge configuration, the underlying mechanism, and the potential role of the surface-anisotropy arising from the thin-film geometry.

\subsection*{Discussion}

The single-phase \AVO~thin films demonstrate strong out-of-plane XRD intensity (Fig.\ref{Fig2}(c)),  sharp interfaces (Fig.\ref{Fig3}), and spatially uniform composition distribution (Fig.S3). However, in contrast to binary V--O films (Fig.\ref{Fig1}(b)), at room temperature, they do not exhibit distinct features in terms of Laue oscillations and RSM peaks (Fig.S1 and S4). We hypothesize that the combined effects of twin domains and charge orderings on macroscopic uniformity play a role. The cubic lattice twinning along the (111) direction produces two-fold domains, while the cubic-to-rhombohedral structural transition produces three-fold domains with varying long-axis orientations (Fig.\ref{Fig4}(c)). Given the limited understanding of the (1/3, 1/3, 1/3)$_{pc}$ superlattice, the existence and structure of the resulting domains also remain unresolved. In our STEM analysis, only the twin domain is clearly resolved. Considering the restricted thermal energy and small thickness inherent to thin-film synthesis, we surmise that the charge-ordering domains do not grow to a sufficiently large size. Indeed, we observed a marked improvement in the \AVO~RSM characteristics at 750 K in terms of intensity and definition (Fig.S4). This may also reflect quantitative differences in the electrical properties of the thin-film and bulk \AVO~(Fig.\ref{Fig3}(e)).

Finally, we note that during the two-step synthesis of \AVO, additional peaks appear at lower 2$\theta$ than the \AVO~peak under slightly oxygen-deficient \Poa~conditions (Fig.\ref{Fig2}(c)). The same behavior was also observed when using the V precursor (Fig.S2). This phase cannot be synthesized as a single-phase material, and its limited intensity makes unambiguous characterization difficult. The out-of-plane XRD peaks of this phase show a periodicity similar to that of \AVO~((222)$_{pc}$ and (333)$_{pc}$). Based on the structural similarity and the oxygen-deficient conditions, we hypothesize that this is an Al(V$_{2-x}$,Al$_x$)O$_4$ compound with a cubic spinel structure. The end member Al$_2$V$^{2+}$O$_4$ has previously been discussed, but a single-phase material has not yet been realized~\cite{camara2018dellagiustaite}. Substitution of Al$^{3+}$ for V$^{3+}$ could be induced under oxygen-deficient conditions, lowering the average V oxidation state of the film.

\section*{Conclusion}

This study highlights a unique advantage \textit{in-situ} interfacial topotaxy offers for fabricating thin films of a complex suboxide, \AVO. By balancing the dynamic processes of oxidation and interfacial reaction, a two-step protocol was established for high-quality epitaxial \AVO/\AO~(0001) thin films. In addition to featuring well-defined epitaxial relationships and a sharp heterointerface, the \AVO~thin film reproduces the structural and electrical properties previously reported for its polycrystalline form. As demonstrated, interfacial topotaxy and \textit{in-situ} oxidation control can work synergistically to create artificial heterostructures with finely tuned charge states. We therefore propose that the \textit{in-situ} interfacial topotaxy technique warrants further investigation and may find widespread utility in the synthesis of complex refractory ternary oxides, and beyond.

\section*{Methods}
\textit{Molecular beam epitaxy}: The details of the CO$_2$~laser-assisted oxide molecular beam epitaxy are comparable to those described in our previous studies~\cite{kim2025high,glotzer2026thermally}. Sample growth was carried out in an UHV chamber  with a base pressure below 1$\cdot$10$^{-9}$ mbar. A custom optical setup incorporating a CO$_2$ laser was used to heat the substrate. The substrate temperature was measured from the back side using a pyrometer, whose emissivity was calibrated by melting an \AO~(0001) substrate at approximately 2040~\dC. Molecular oxygen gas was supplied through two leak valves connected in series. The oxygen supply was quantified by a local pressure between the two valves (\Po), measured by a capacitance manometer. The \Po~and the stabilized growth chamber pressure were found to be proportional. In this study, \Po~values of 0 and 0.4 mbar corresponded to growth chamber pressures of $<$1$\cdot$10$^{-9}$~and~1$\cdot$10$^{-6}$ mbar, respectively. Elemental vanadium was supplied using an electron beam evaporator, and its flux was measured with a quartz crystal microbalance. For all growths, the V flux was maintained at $\sim$0.09 \AA/s , and the \AO~(0001) substrates were pre-annealed at 1500~\dC~under UHV conditions \textit{in situ}. The pre-annealing, growth, and post-annealing durations were 10, 60, and 30 minutes, respectively.

\textit{X-ray diffraction}: X-ray diffraction measurements were conducted using a Rigaku SmartLab diffractometer equipped with a Cu-K$\alpha_{1}$ source. Reciprocal space mapping data were collected using a Rigaku HyPix-3000 1D detector. Temperature-dependent measurements were performed on \AVO~samples mounted on an Anton Paar DCS 500 stage under a background pressure below 3$\cdot$10$^{-1}$ Torr. To accurately determine the peak positions over a wide temperature range, we used the separation between the \AO~(0006) and (00012) peaks as a calibration reference. Since the peak separation was largely invariant with respect to the $\omega$ offset, the $\omega$ offset at each temperature was adjusted to satisfy the self-consistency condition imposed by the peak separation. The linear thermal expansion of the \AVO~and \AO~lattices were fitted by $d$~$\approx$~1.4043 + 1.6338$\cdot$10$^{-5}\cdot$$T$ and 1.2957 + 0.7324$\cdot$10$^{-5}\cdot$$T$, respectively. 

\textit{Electrical transport}: Temperature-dependent resistivities were measured using a Quantum Design DynaCool system. Thin-film samples, cleaved to $\sim$2$\times$2 mm in size, were wire-bonded in a Van der Pauw geometry. Resistivity was calculated from the measured sheet resistance and film thickness determined by XRD.

\textit{Electron Microscopy}: Cross-sectional TEM and STEM were performed on a JEOL NEOARM aberration corrected microscope operated at 200~kV. Images were collected using Annular Dark Field and Bright Field detectors. Energy Dispersive X-ray Spectroscopy (EDS) data were collected using an Oxford Instruments AZtecEnergy EDS system coupled to two JEOL 100~mm$^2$ silicon drift detectors. Lamellae were prepared using a Thermo Fisher Helios NanoLab 650 Dual Beam microscope with a Ga liquid metal ion source, operating at 30 and 5kV for thinning and final polishing, respectively.

\section*{Acknowledgments}
The authors thank Seokhwan Yun and Darrell G. Schlom for their valuable comments. We acknowledge funding provided by the Institute for Quantum Information and Matter, a NSF Physics Frontiers Center (NSF Grant No. PHY-2317110). This research was supported by Basic Science Research Program through the National Research Foundation of Korea (NRF) funded by the Ministry of Education (RS-2024-00413513). S. G. are partially supported by the National Science Foundation Graduate Research Fellowship under Grant No. 2139433. Atomic force microscopy data were collected at the Molecular Materials Research Center in the Beckman Institute of the California Institute of Technology. Electron microscopy was carried out in the Nanomaterials Characterization Facility at the University of New Mexico, a facility that is supported by the State of New Mexico, the National Science Foundation and the National Aeronautics and Space Administration. The acquisition of the JEOL NEOARM AC-S/TEM at the University of New Mexico was supported by NSF grant DMR-1828731 and NASA Emerging Worlds grant 80NSSC21K1757.

\clearpage

\bibliographystyle{unsrt}
\bibliography{ref.bib}

\clearpage

\renewcommand{\thefigure}{S\arabic{figure}}
\renewcommand{\thetable}{S\arabic{table}}

\setcounter{figure}{0}

\begin{table*}[h]
\centering
\caption{
	\label{tableS1} 
	Crystal structures of V--O and \AVO
      }
        \begin{tabular}{c|c|ccc|ccc}\hline\hline
          Compounds & Space group & a & b & c (\AA) & $\alpha$ & $\beta$ & $\gamma$ ($^\circ$) \\\hline
          \AO & $R\overline{3}c$ & 4.76 & 4.76 & 13.00 & 90 & 90 & 120 \\\hline
          V & $Im\overline{3}m$ & 3.03 & 3.03 & 3.03 & 90 & 90 & 90 \\
          V$_4$O & $I4/mmm$ & 2.98 & 2.98 &  3.37 & 90 & 90 & 90 \\
          V$_2$O & $P2_{1}/c$ & 9.55 & 2.92 & 7.76 & 90 & 90.32 & 90 \\
          VO & $Fm\overline{3}m$ & 4.08 & 4.08 & 4.08 & 90 & 90 & 90 \\
          V$_2$O$_3$ & $R\overline{3}c$ & 4.95 & 4.95 & 14.00 & 90 & 90 & 120 \\\hline
          $c$-AlV$_2$O$_4$ & $Fd\overline{3}m$ & 8.15 & 8.15 & 8.15 & 90 & 90 & 90 \\
          $r$-AlV$_2$O$_4$ & $R3m$ & 5.76 & 5.76 & 28.86 & 90 & 90 & 120 \\
    \hline\hline
  \end{tabular}
\end{table*}

\begin{table*}[h]
\centering
\caption{
	\label{tableS2} 
	Epitaxial relationships of V--O and \AVO~with respect to the \AO~(0001) substrate
      }
  \begin{tabular}{c|ccc}\hline\hline
          Compounds & ~~Out-of-plane~~ & ~~In-plane~~ \\\hline
          \AO & (0001) & [11$\overline{2}$0] & [1$\overline{1}$00] \\\hline
          V & (110)  &  [1$\overline{1}$2] & \\
          V$_4$O & (101) & [$\overline{1}$01] & \\
          V$_2$O & (203) &   & [010]\\
          VO & (111) & [11$\overline{2}$] & \\
          V$_2$O$_3$ & (0001) & [11$\overline{2}$0] & \\\hline
          $c$-AlV$_2$O$_4$ & (111) & [11$\overline{2}$] & \\
          $r$-AlV$_2$O$_4$ & (1$\overline{1}$2) & [$\overline{1}$11] & \\
    \hline\hline
  \end{tabular}
\end{table*}

\begin{figure*}[t]
  \centering
   \includegraphics[width=170mm]{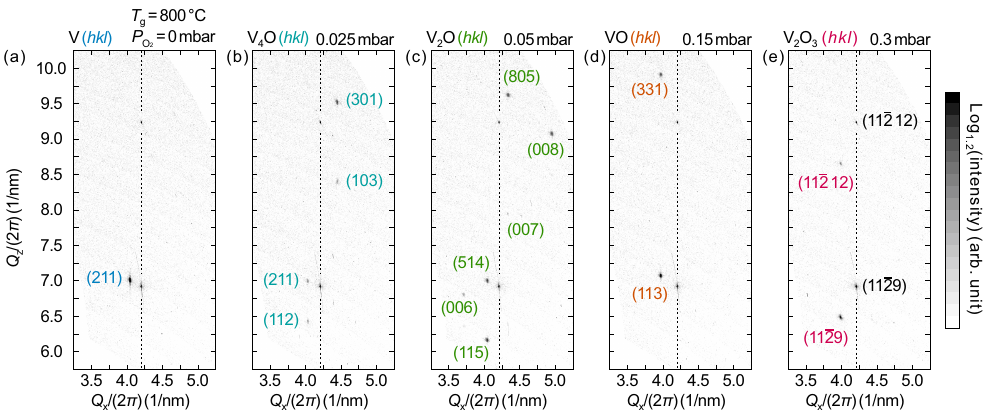}
   \caption{XRD reciprocal space mappings of the five single-phase V (a), V$_4$O (b), V$_2$O (c), VO (d), and V$_2$O$_3$ (e) films grown on \AO~(0001) substrates. The samples are identical to those shown in Figure 1(b).}
   \label{FigS1}
\end{figure*}

\begin{figure*}[t]
  \centering
   \includegraphics[width=170mm]{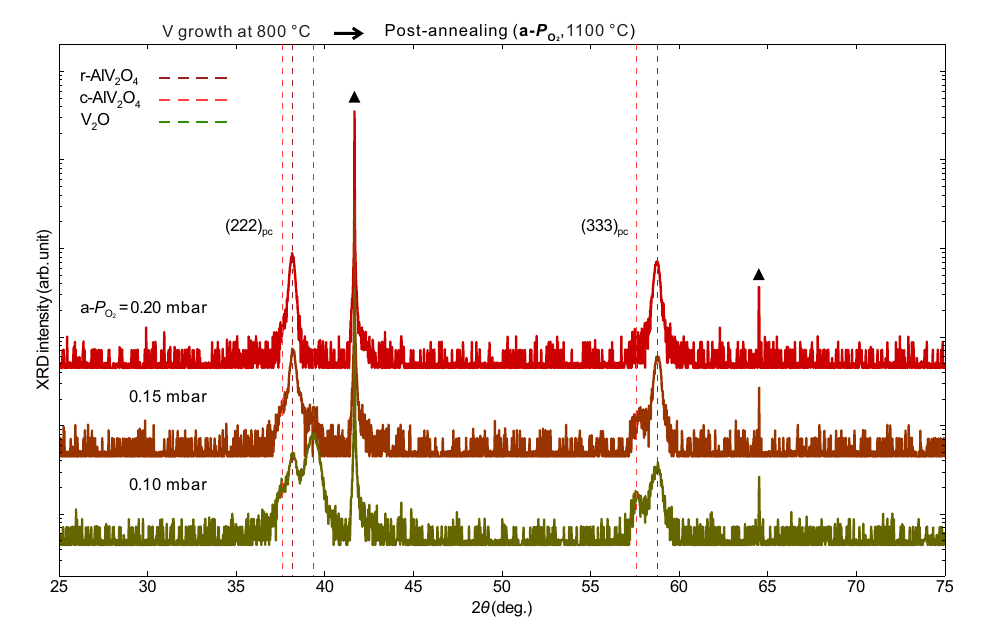}
   \caption{Wide-range XRD 2$\theta$--$\theta$ plots of the 1100~$^\circ$C-post-annealed samples from V precursors. The samples correspond to (\Poa,~\Pog) = (0.1, 0), (0.15, 0), (0.2, 0) points in Figure 2(a).}
   \label{FigS2}
\end{figure*}

\begin{figure*}[t]
  \centering
   \includegraphics[width=170mm]{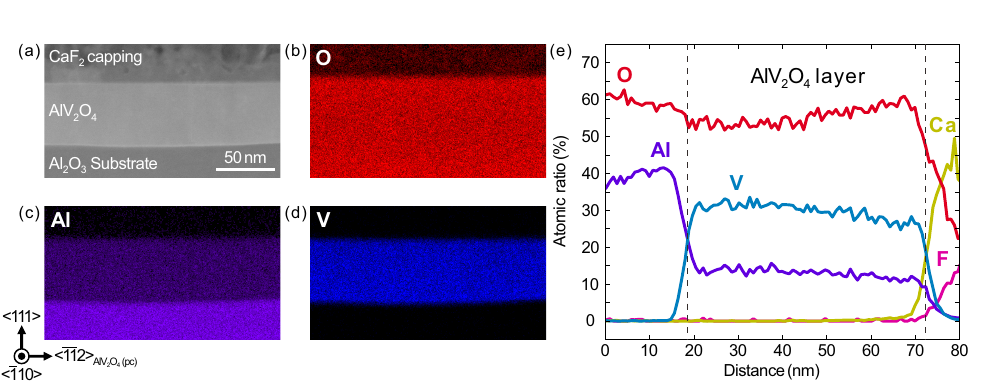}
   \caption{Composition analysis at the \AVO/\AO~heterointerface. (a--d) HAADF-STEM (a) and TEM energy dispersive spectroscopy mappings of O (b), Al (c), and V (d) elements, taken from the \AO~[1$\overline{1}$00] zone axis. Depth profiles of relative elemental amounts.}
   \label{FigS3}
\end{figure*}

\begin{figure*}[t]
  \centering
   \includegraphics[width=85mm]{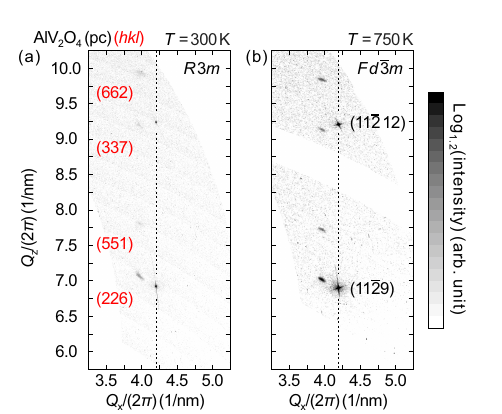}
   \caption{XRD reciprocal space mappings of a \AVO/\AO~(0001) film at $T$ = 300 K (a) and 750 K (b).}
   \label{FigS4}
\end{figure*}

\end{document}